\documentclass[journal]{IEEEtran}
\usepackage{epsfig,rotating,setspace,latexsym,amsmath,epsf,amssymb,amsfonts,bm,theorem,booktabs,subfigure,epstopdf}
\usepackage{cite,authblk,hyperref}
\usepackage{bbm}
\usepackage{amsmath}
\usepackage{amssymb}
\usepackage{bm}
\usepackage{makecell}
\usepackage{graphicx}
\usepackage{flushend}
\usepackage{algorithm}
\usepackage{algpseudocode}%
\usepackage{color}

\definecolor{commentcolor}{RGB}{110,154,155}

\IEEEoverridecommandlockouts
\allowdisplaybreaks

\begin{document}
	\bstctlcite{IEEEexample:BSTcontrol} 
	
	\title{MCANet: A Coherent Multimodal Collaborative Attention Network for Advanced Modulation Recognition in Adverse Noisy Environments}
	
	\author[1]{Wangye Jiang}
	\author[2]{Haoming Yang}
    \author[1]{Xinyu Lu}
    \author[1]{Mingyuan Wang}
    \author[1]{Huimei Sun}
    \author[1]{Jingya Zhang}

\affil[1]{\normalsize School of Electronic and Information Engineering, Suzhou University of Technology, Suzhou 215500, China \\ \texttt{zhangjy0611@163.com}} 
\affil[2]{\normalsize School of Software Engineering, Jinling Institute of Technology, Nanjing 211169, China}
\maketitle

\begin{abstract}
As wireless communication systems evolve, automatic modulation recognition (AMR) plays a key role in improving spectrum efficiency, especially in cognitive radio systems. Traditional AMR methods face challenges in complex, noisy environments, particularly in low signal-to-noise ratio (SNR) conditions. This paper introduces MCANet (Multimodal Collaborative Attention Network), a multimodal deep learning framework designed to address these challenges. MCANet employs refined feature extraction and global modeling to support its fusion strategy.Experimental results across multiple benchmark datasets show that MCANet outperforms mainstream AMR models, offering better robustness in low-SNR conditions.
\end{abstract}

\begin{IEEEkeywords}
	Automatic modulation recognition, wireless signal classification, contrastive learning, deep learning, self-supervised learning, spectrum awareness. 
\end{IEEEkeywords}

\section{Introduction}\label{sec:Introduction}
\emph AMR of signals, as a signal processing technology used to identify unknown or partially known communication signal transmission parameters, plays a crucial role in cognitive radio applications \cite{1}, particularly in satellite communications \cite{2}. With the advancement of software-defined radio and cognitive radio in civilian wireless communications, AMR has become the core enabling technology for cognitive receivers. It not only alleviates the issues of communication resource consumption and reduced system efficiency caused by the exchange of transmission parameters over dedicated channels, but also facilitates dynamic, demand-driven wireless resource management in cognitive radio, helping to ease the problem of spectrum congestion. This is essential for enhancing the performance and resource utilization of wireless communication systems \cite{3}.

Traditional AMR methods can be broadly categorized into likelihood-based AMR (LB-AMR) \cite{4}, \cite{5} and feature-based AMR (FB-AMR) \cite{6}, \cite{7}. The LB-AMR method makes classification decisions using likelihood ratio functions. Although it can achieve higher classification accuracy with increased complexity, its poor real-time performance has made it inadequate for meeting the timeliness and rapid response requirements of modern wireless communication systems. For example, the Expectation-Maximization (EM) method \cite{8}, \cite{9}, which relies on maximizing the likelihood function to estimate parameters, has strong classification ability. However, due to its complex and time-consuming computation process, it is challenging to apply effectively in contemporary communication environments.

On the other hand, FB-AMR methods primarily rely on expert knowledge to manually select features, using simple classifiers such as Support Vector Machines (SVM) \cite{10,11,12}, Random Forest (RF) \cite{13,14}, and Decision Trees (DT) \cite{15,16,17} for supervised learning. Although these methods are highly practical, their heavy dependence on expert priors and fixed feature sets often leads to suboptimal solutions. This lack of flexibility and scalability makes them ineffective when confronted with new modulation schemes, as they fail to adapt appropriately. 

In recent years, deep learning-based AMR (DL-AMR) has brought revolutionary changes, significantly improving the accuracy and efficiency of related tasks. By relying on self-learning through neural networks, DL-AMR can also reduce the need for human resources to a certain extent. O'Shea and West proposed an open-source synthetic radio machine learning dataset built on GNU Radio, which incorporates real-world channel effects \cite{18}, considering various factors such as Doppler shifts, multipath fading, and carrier frequency offset. This dataset addresses the lack of standardized benchmark datasets in the radio field, adapting complex baseband signals into formats compatible with machine learning frameworks. It also expands the scope of radio machine learning tasks, providing example classification models and dataset acquisition channels, thus establishing performance comparison benchmarks for the field. At the same time, neural image classification technologies have undergone rapid development. Since the introduction of AlexNet \cite{19}, various neural networks for image recognition have emerged, driving the expansion of AMR-related solutions. Zhao compared the performance of CNN (Convolutional Neural Network) and ResNet in AMR tasks, analyzing the underlying mechanisms \cite{20}. Zhao et al. were the first to apply a mask autoencoder-based Vision Transformer (ViT-MAE) as a foundational model for AMR, setting a new standard for robustness and accuracy in transformer-based modulation classification tasks \cite{21}. Faysal et al. proposed NMformer \cite{22}, innovatively transforming multi-SNR modulation signals into RGB constellation diagrams. By utilizing a Vision Transformer to capture global features and fine-tuning with small samples, the model improves classification performance for low-SNR and out-of-distribution data, addressing the core issue of noisy modulation classification without the need for extra denoising. However, while these image-based approaches have significantly advanced AMR performance, they also expose fundamental limitations. From the perspective of image modality processing logic, current mainstream solutions rely on CNNs for feature extraction. The core advantage of CNNs lies in their ability to model spatial structures, allowing them to precisely capture local spatial features—such as spectral textures and frequency domain energy distributions—in the image-like representations formed by time-frequency transformations of the signal. This is also the core reason why CNNs perform exceptionally well in image processing. Yet, the intrinsic nature of a signal is dynamic temporal flow, and the image modality inevitably loses temporal dynamics during the conversion of the signal into a static spatial form. Furthermore, CNNs lack the ability to model temporal dependencies, making it impossible to recover these dynamic features. This ultimately limits classification accuracy when processing modulated signals with complex temporal variations, as the model relies solely on static spatial features. On the other hand, time-series data processing technologies, particularly for signal processing and communications domain analysis, have also seen rapid development. In this context, Long Short-Term Memory networks (LSTM) and Recurrent Neural Networks (RNN), as mainstream models for time-series data processing, have been widely applied to AMR-related tasks, giving rise to several DL-AMR methods \cite{23,24,25,26}. However, with the increasing demand for long-sequence modeling capabilities and parallel efficiency, the "vanishing/exploding gradient" problem in RNNs and the "serial computation bottleneck" that still exists in LSTM, despite gating mechanisms, have become increasingly prominent, making them difficult to adapt to more complex time-series scenarios. This technical evolution has led to the recent emergence of Transformer models. As a high-quality model for time-series tasks in recent years, Transformer’s unique self-attention mechanism has enabled it to achieve strong results across various domains. Not only has it shown its distinct role in time-series tasks \cite{27,28,29}, but it has also made significant breakthroughs in computer vision (CV) \cite{30,31} and natural language processing (NLP) \cite{32,33}. Consequently, an increasing number of researchers have begun applying Transformer to AMR tasks. Cai et al. \cite{34} were the first to apply Transformer networks to AMR problems. By effectively utilizing global information from sample sequences, they significantly improved classification accuracy under low SNR conditions. Compared to other deep models, this approach required fewer training parameters and was more suited for practical applications. Kong et al. proposed a semi-supervised contrastive learning framework based on Transformer \cite{35}, integrating convolutional embeddings into the encoder. This framework utilizes unlabeled samples for self-supervised contrastive pre-training and applies time-warping for data augmentation. The framework improves classification accuracy by fine-tuning the pre-trained encoder and the randomly initialized classifier, even with limited labeled samples. Furthermore, CNNs have also played an active role in time-series tasks. Tu et al. proposed a CNN-based model that simultaneously employs complex-valued networks for AMR \cite{36}, making full use of the information in complex signals while clarifying the complexity and optimization direction. However, the limitations of time-series modalities are equally significant. On one hand, RNNs and LSTMs tend to suffer from vanishing or exploding gradient issues when processing long sequence signals, making it difficult to effectively model long-term temporal dependencies. On the other hand, time-series modalities focus only on the local dynamic features of a one-dimensional temporal flow and lack the ability to integrate global signal information. Especially in complex noisy environments, local temporal features are easily interfered with by noise, and the absence of global information further amplifies the negative impact of noise, leading to a significant reduction in classification robustness. Furthermore, even Transformer models, which have the capability for global modeling, still fall short in capturing the fine temporal details of the signal compared to RNNs/LSTMs. Additionally, Transformer models require a large amount of labeled data and show insufficient robustness when faced with noisy or out-of-distribution data, making them less optimal for time-series modalities.

Fundamentally, the core requirement of automatic modulation format classification is to comprehensively capture the multi-dimensional intrinsic features of the signal. The time-domain dynamics and frequency-domain spatial characteristics of the signal are not contradictory but complementary key dimensions. Time-domain features reflect the dynamic change patterns of the signal, while frequency-domain image features reflect the signal's global energy distribution and spatial correlations. A single modality can only cover one of these dimensions and fails to provide a complete representation of the signal's features. Therefore, in low SNR and complex modulation scenarios, the limitations of a single modality are further magnified, making it difficult to meet the practical application requirements of classification performance. Based on this, multimodal learning technology, which integrates both image and time-series modalities, has become the core research trend for overcoming the limitations of single-modal approaches. The fusion of modal data and the construction of complex models can not only enhance the adaptability to complex signal environments but also leverage the complementary nature of multimodal features to counteract noise interference. Ultimately, this significantly improves the automatic modulation format classification accuracy in low-SNR and out-of-distribution data scenarios, providing a more robust and comprehensive technical solution for AMR tasks.

Similar multimodal feature fusion experiments have been conducted by previous researchers. Samarkandi et al. proposed a deep learning-based CNN automatic modulation classification algorithm \cite{37}, which optimizes classification performance in communication signal modulation by using joint concatenated images of constellation diagrams, ambiguity functions (AF), and eye diagrams as inputs. The classification performance of the combined constellation diagram and eye diagram outperforms their individual use, while the combination of AF and eye diagram enhances performance under low SNR, providing an efficient solution for spectrum sensing. Zhao and Luo \cite{38} proposed a novel automatic modulation classification scheme based on transfer learning, which converts multimodal signals into I/Q spectra and constellation diagrams to adapt to visual models. This approach outperforms traditional methods on the RadioML dataset, with experimental verification showing that multimodal fusion significantly enhances the model’s robustness in classifying modulated signals in noisy environments. Zhuang et al. \cite{39} combined time-domain signals with continuous wavelet transform (CWT) based on complex Morlet wavelets. This innovative method enhanced the model’s anti-interference capability, thus improving the accuracy of feature extraction in composite material damage detection. Han and Yang \cite{40} proposed an automatic modulation classification method based on an attention mechanism by fusing I/Q time-series data with constellation diagrams, SPWVD time-frequency diagrams, and SCF, using convolutional autoencoders to denoise image features and a self-attention multilayer perceptron (MLP) for fusion. Experimental results demonstrated that the multimodal fusion-based model exhibited superior robustness.

In general, although the aforementioned multimodal fusion methods effectively enrich the input features, they still face certain limitations in low SNR environments. Due to noise interference in low SNR conditions, the useful information in the signal is often obscured, causing independently extracted modal features to fail in capturing multi-modal joint information, which is crucial for the success of classification tasks. Feature extraction from a single modality in such environments cannot comprehensively reflect the signal characteristics. Therefore, joint mining of multi-modal information is particularly important. Only by fully exploiting the complementarity and spatial features between modalities can classification accuracy be improved. Traditional fully connected layer concatenation methods struggle to capture the complex nonlinear relationships between modalities, and hybrid models still have limitations in capturing global feature dependencies and efficiently fusing multimodal features. Thus, existing multimodal fusion methods face challenges in low SNR environments, including incomplete signal feature extraction, insufficient information fusion, and severe noise interference, requiring more effective multi-modal fusion mechanisms to enhance classification accuracy and robustness.

To address the aforementioned challenges, this paper proposes a DL-AMR hybrid framework, named the MCANet. This framework leverages multimodal feature fusion to achieve comprehensive signal representation and improved classification performance in noisy environments. In terms of time-frequency domain feature selection, we employ a wavelet transform-based feature processing approach, which decomposes the signal into low-frequency and high-frequency components. The low-frequency portion is enhanced to highlight the main information, while the high-frequency part is attenuated through learnable weights to reduce noise interference, thus yielding more robust temporal input features.

In the image branch, since the constellation diagram and eye diagram each provide spatial and temporal features of the signal, respectively, their multimodal fusion enhances the model’s robustness in low-SNR environments \cite{37}. In this paper, constellation and eye diagrams are used as inputs in multi-modal modeling, processed separately through an improved ResNet \cite{41} dual-branch network to extract spatial modal features and perform deep representation learning. Additionally, in the time-series branch, we employ a Transformer encoder to globally model the wavelet-transformed embedded sequence features. This design allows for the extraction of global context information while preserving local details. As the image branch uses the ResNet network for spatial feature extraction, and the time-series branch employs the Transformer encoder for global modeling of the wavelet-transformed features, this approach effectively balances local details and global context information. As a result, the model’s feature representation capability is enhanced, while simultaneously boosting the model's robustness to complex signals.

In the feature embedding and fusion stages, we design a Spatial-Channel Attention with Position Encoding (SCAPE) module, based on the Dual Coordinate Attention Feature Extraction (DCAFE) module from the Flora-NET model proposed by Gupta and Tripathi \cite{42}. This module maps temporal features to the image feature space during the fusion stage, achieving efficient multimodal feature fusion through projection convolution. It innovatively introduces spatial geometric encoding and extends the original attention mechanism’s coordinate dimension to the spatial dimension, forming a multi-dimensional attention collaborative mechanism. Through precise spatial-semantic mapping, the module performs deep semantic adaptive fusion of multi-modal features, providing a robust feature representation foundation for complex signal recognition tasks.

The key contributions of this research can be summarized as follows:

1. A dual-branch collaborative framework has been constructed, integrating ResNet for local feature extraction and Transformer for global dependency modeling. The ResNet branch focuses on fine extraction of image spatial local features, while the Transformer branch is responsible for deep modeling of temporal global dependencies, achieving an efficient balance and complementarity between local details and global context.

2. An attention mechanism has been developed, focusing on three dimensions to dynamically optimize feature weight distribution through precise spatial-semantic mapping. This design enhances the model’s ability to adaptively emphasize informative features while suppressing redundant or noisy signals, contributing to more robust and discriminative feature representations.

3. Integration of wavelet frequency-domain optimization with the deep learning framework enables adaptive processing of low-frequency enhancement and high-frequency suppression, improving the model’s robustness in complex environments.

\section{Signal Model} \label{sec:SignalModel}
In the traditional framework of wireless communication systems, the transmitted wireless signal in real-world environments can typically be accurately modeled by the convolution of the channel impulse response with the normalized wireless signal, with the addition of uncorrelated Gaussian noise. Mathematically, this relationship can be expressed as:
\begin{align}
	X(n) = H(n) * S(n) + W(n)
\end{align}

where $H(n)$ represents the channel impulse response, $S(n)$ denotes the normalized wireless signal, the symbol “*” represents the convolution operation, and $W(n)$ is the uncorrelated additive Gaussian noise process, which is a common noise component in simulation and modeling frameworks. The intensity of the noise is typically quantified by the SNR, an important metric for assessing the quality of the wireless signal.

The goal of AMR is to identify the modulation type of the signal $X(n)$ based on the observed $rIQ(n) $, where the signal is affected by both channel effects and noise.  For convenience in processing, $X(n)$ in this paper is pre-processed and converted into in-phase (I) and quadrature (Q) components. At this point, the received signal can be represented in complex form as:
\begin{align}
	r_{IQ}(n) = X_I(n) + jX_Q(n)
\end{align}

where $X_I()$ and $X_Q()$ represent the in-phase and quadrature components of the signal, respectively, $l$ typically represents the index of discrete time, and $j$ is the imaginary unit. This I/Q representation effectively maps the signal’s time-domain information and modulation characteristics onto the complex plane, facilitating subsequent signal analysis and modulation classification.

To reveal the modulation details and time-domain characteristics of the signal, this paper transforms the IQ component signal into constellation diagrams, eye diagrams, and time-frequency features extracted through wavelet transform. Unlike traditional time-frequency transformation methods, the constellation diagram provides an intuitive representation of the signal's distribution in the complex plane, while the eye diagram demonstrates the signal's time-domain variations by overlaying waveforms of multiple symbol periods. Additionally, the wavelet transform offers a detailed decomposition of the signal at different scales, enabling the simultaneous capture of both time-domain and frequency-domain features of the signal. Therefore, by combining these three features, the robustness of signal features is enhanced, enabling a more comprehensive analysis and identification of different modulation schemes. The principles behind these three features will be introduced in detail below.

\subsection{Constellation Diagram}
The generation of the constellation diagram is based on the I and Q components of the signal, which are mapped onto the complex plane to form a scatter plot. Specifically, this can be expressed in set notation as:
\begin{align}
	\mathcal{S}_{\text{cons}} [i]= \left\{ (X_I(n), X_Q(n)) \mid n = 1, 2, \dots, N \right\}
\end{align}

where $\mathcal{S}_{\text{cons}}[i]$ represents the constellation diagram trajectory set for the $i-th$ signal and $N$ represents the total number of symbols in the signal. In this way, the constellation diagram visually presents the different modulation schemes of the signal on the complex plane. The geometric structure of the diagram directly reflects the modulation format of the signal. This visual feature 
provides an intuitive basis for modulation type recognition and signal quality analysis, facilitating the classification of signals by the model.
\subsection{Eye Diagram}

\begin{figure}[H]
	\centering
	\includegraphics[width=0.8\columnwidth]{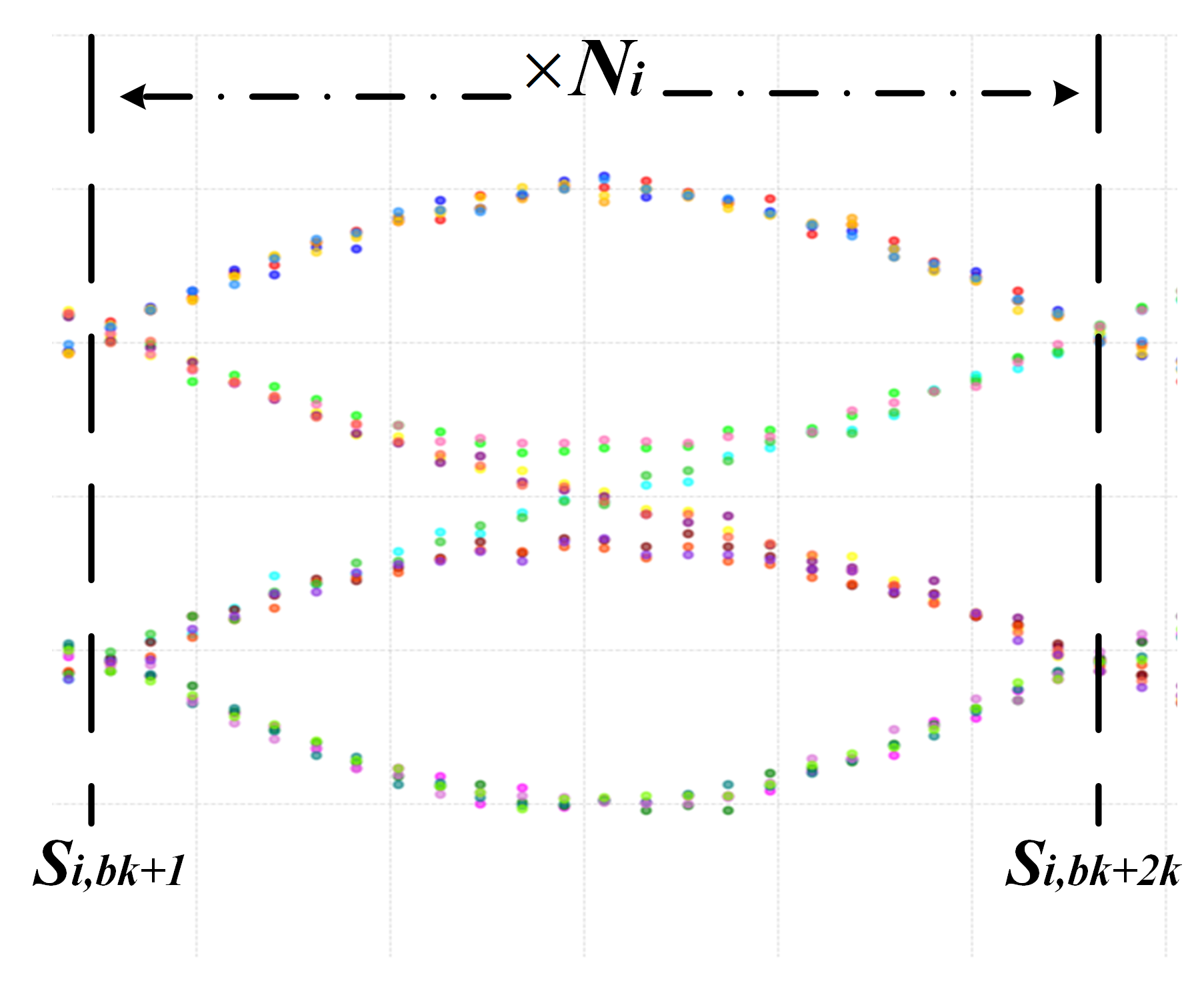}
	\caption{Sampling Instance Eye Diagram} 
	\label{Sampling Instance Eye Diagram}
\end{figure}

The principle of the eye diagram is based on the superposition of multiple symbol periods of the signal, allowing it to clearly display the waveform of the signal at the receiver and reveal signal distortions and inter-symbol interference. The generation process can be formulated as:

\begin{align}
	\mathcal{S}_{\text{eye}}[i] = \bigcup_{b=0}^{N_i-1} \left\{ s_{i,bk+1}, \dots, s_{i,bk+2k} \right\}, \quad N_i = \left\lfloor \frac{|L_i|}{k} \right\rfloor - 1
\end{align}

where $\mathcal{S}_{\text{eye}}[i]$ represents the eye diagram trajectory set for the $i-th$ signal, $s_{i,k}$ is the $k-th$ sampling point of the $i-th$ signal sequence, $Li$ is the total length of the $i-th$ signal sequence, $K$ is the number of sampling points per symbol, and $Ni$ is the number of eye diagram trajectories that can be extracted from the $i-th$ signal. The segment $bK+1:bK+2K$ represents the trajectory segment starting from the $bK+1-th$ sampling point with a length of $2K$, as shown in Fig.~\ref{Sampling Instance Eye Diagram}.

\subsection{Wavelet Transform}
Finally, to further enhance the time-frequency features of the signal, this paper introduces the wavelet transform. Compared to traditional Fourier transforms, wavelet transforms can simultaneously capture variations of the signal in both the time and frequency domains, making them particularly suitable for the analysis of non-stationary signals. Through wavelet transform, the signal is decomposed into components at different frequencies, thus obtaining multi-scale features of the signal. Specifically, the mathematical expression for the wavelet transform of a signal $r_{IQ} $ is:

\begin{align}
	w(j, k) = \sum_l r_{IQ}[n] \cdot \varphi^*_{jk}[n]
\end{align}

where $\varphi[]$ is the wavelet basis function, $j$ is the scale parameter, which controls the frequency analysis range, and $k$ is the translation parameter, which controls the time analysis position. In this paper, the Daubechies wavelet (db4) is used to perform a two-level wavelet decomposition of the signal, yielding the low-frequency component $cA2$ and two high-frequency components $cD2$ and $cD1$. These components, $cA2$,$cD2$,$cD1$, are the multi-scale features extracted by the DWT, which represent discrete samples of $w(j,k) $ at these specific scales. They facilitate the extraction of both local and global information while filtering out high-frequency noise. 

\section{Methodology} \label{sec:ssl}
\begin{figure*}[tb!]
	\centering
	\includegraphics[width=0.8\textwidth]{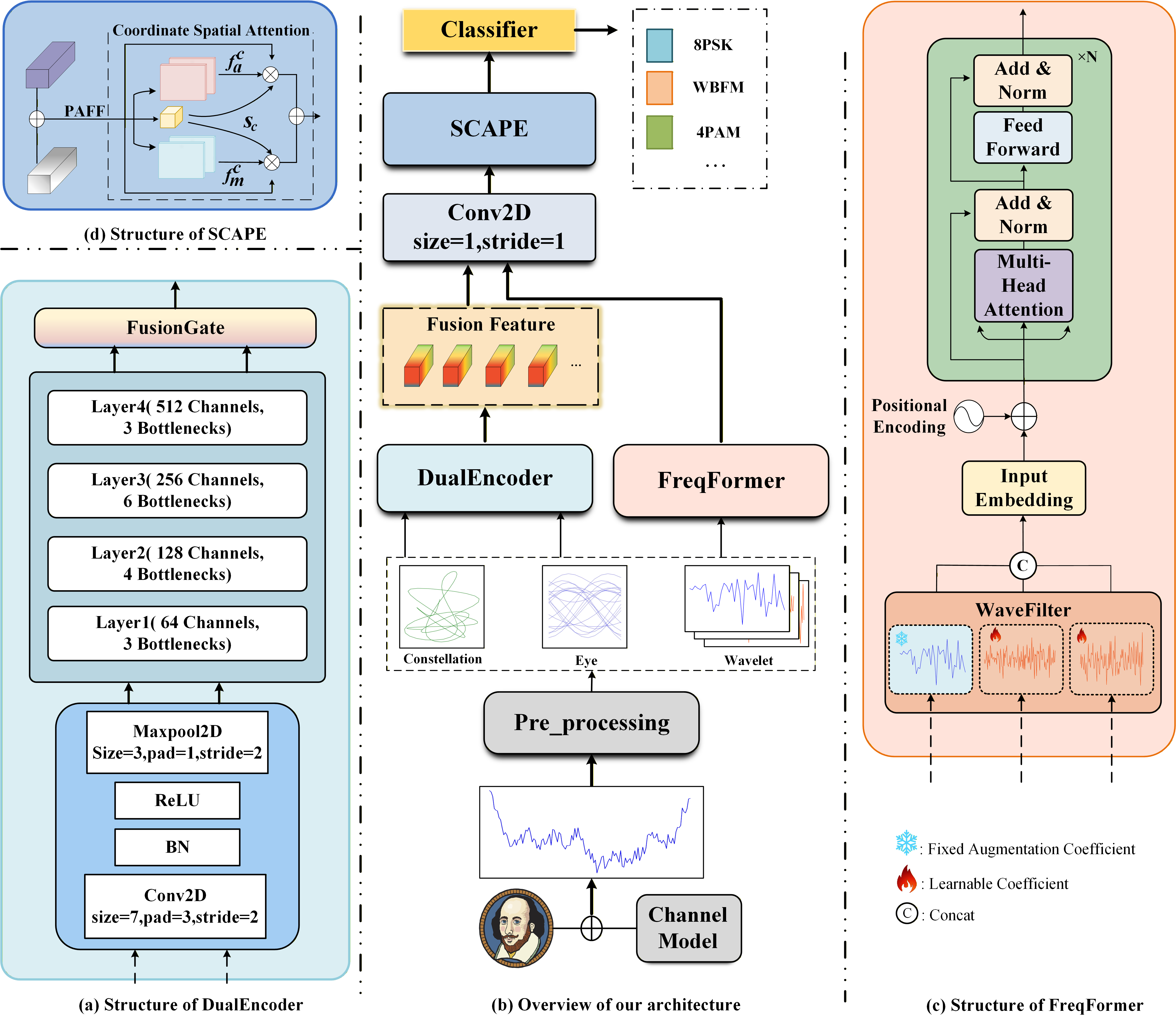} 
	\caption{The overall network architecture} 
        	\label{The Overall Network Architecture}
\end{figure*}

The overall architecture of the proposed framework is illustrated in Fig.~\ref{The Overall Network Architecture}, comprising three main components: the time-series feature processing module (herein referred to as Freqformer), the image feature processing module (DualEncoder), and the feature fusion module (termed SCAPE). The modulation recognition process described in this paper can be summarized as follows: In the preprocessing stage, the IQ signal undergoes multimodal transformation to generate constellation diagrams, eye diagrams, and wavelet features. Subsequently, Freqformer and DualEncoder process these multimodal features in parallel. DualEncoder utilizes the improved dual-branch ResNet50 network to dynamically extract deep spatial features from the constellation diagram and eye diagram, while Freqformer performs dynamic filtering of the wavelet features. Fixed low-frequency enhancement coefficients and learnable high-frequency attenuation coefficients are applied to highlight the primary signal components and suppress noise. Next, the time-series features pass through the sequence embedding layer and the Transformer encoder to extract temporal features. Finally, the SCAPE concatenates and linearly fuses the time-series and image features. After adaptive fusion, the fused features are input into a fully connected classifier for modulation recognition.
The following sections will provide a detailed description of the design of these modules.
\subsection{DualEncoder}
In DualEncoder, the constellation diagram and eye diagram serve as two primary signal representations, each providing rich information about the signal in the frequency and time domains, respectively. However, the information from these two modalities is typically presented in different formats, leading to challenges in modal alignment and feature fusion dimension mismatch during the fusion process. Traditional feature fusion methods often rely on simple concatenation or weighted averaging, which fail to effectively address the differences between modalities in both spatial and temporal domains. Additionally, they lack adaptive fusion strategies, resulting in information loss or redundancy, thereby impacting the accuracy of modulation recognition. To solve this problem, this paper proposes an innovative image branch design specifically for processing both the constellation diagram and eye diagram simultaneously, effectively addressing the alignment and fusion issues of multi-modal features.

The image branch utilizes the ResNet-50 network for deep spatial feature extraction from both the constellation diagram and eye diagram, and employs a learnable weighted fusion mechanism to dynamically adjust the contribution ratio of the two modal features, thereby achieving more precise multi-modal information fusion.

Specifically, the image branch first applies a 1x1 convolution layer to adapt the channels of the constellation diagram and eye diagram, unifying the shape of the two input images to $H \times W \times 64$, preparing them for subsequent convolution operations. The constellation diagram and eye diagram are then separately processed through the ResNet-50 network to extract convolutional features, capturing the signal's spatial features from low to high levels. Through multiple residual blocks, the network effectively extracts multi-scale features from the images, and the use of batch normalization (BN) and ReLU activation functions enhances the non-linear expressiveness of the features.

Next, the features from the two modalities need to be fused. To achieve this, the image branch uses a learnable fusion weight to perform weighted fusion of the features from the two input modalities. Specifically, a sigmoid activation function is employed to learn and adaptively adjust the fusion weights, allowing the fusion ratio between the two feature streams to be dynamically controlled:

\begin{align}
	z = \alpha x + (1 - \alpha) y
\end{align}

where $\alpha$ is the learnable parameter obtained through the sigmoid function, controlling the weighted ratio of the two feature streams, and $x$ and $y$ are the features of the $\mathcal{S}_{\text{cons}}$ and $\mathcal{S}_{\text{eye}}$ processed by the convolutional network. This mechanism ensures that the features from both modalities are appropriately weighted based on the task's requirements. The weighting mechanism allows the network to flexibly adapt to the information fusion needs under different signal conditions, effectively avoiding issues of information redundancy or loss. The fused feature map is then further reduced in dimensionality through a global average pooling layer, generating a fixed-length feature vector, which is sent to the feature fusion module.
\subsection{FreqFormer}
In FreqFormer, this paper proposes a time-series branch design that combines the Transformer encoder and a high-frequency filtering wavelet feature processor, enabling time-series features to better adapt to the multimodal feature fusion in the modulation recognition task.

The processing flow of the time-series branch begins with the preprocessing of the input 282-dimensional wavelet features using the wavelet feature processor. This processor effectively suppresses the impact of noise through low-frequency enhancement and high-frequency adaptive attenuation, preserving the primary signal components. The processed time-series features have a shape of $B \times 282$, and are then passed through a fully connected layer, mapping the 282-dimensional features into a higher-dimensional time-series space, with an output dimension of $B \times 8 \times d\_{\textit{model}}$, where $d\_{\textit{model}}$ =64 represents the embedding dimension of each time step.

Next, the time-series features, enhanced with positional encoding, are fed into the Transformer encoder. The positional encoding introduces relative positional information, enhancing the spatial structure of the time-series features across different time steps.
\begin{align}
	x_{\text{encoded}} = x_{\text{emb}} + pos_{proj}
\end{align}

Here, $x_{\text{emb}}$ represents the time-series embedded features from the fully connected layer, and $pos_{proj}$ is the position encoding learned for each time step $t$.

Subsequently, the feature $x_{\text{encoded}}$, which now includes positional information, is input into the Transformer encoder. The Transformer encoder models the global dependencies between features based on the self-attention mechanism. In this mechanism, the Query ($Q$), Key ($K$), and Value ($V$) matrices are obtained from the encoder input sequence via linear projections, as follows:
\begin{align}
	Q = x_{\text{encoded}} \cdot W_Q
\end{align}
\begin{align}
	K = x_{\text{encoded}} \cdot W_K
\end{align}
\begin{align}
	V = x_{\text{encoded}} \cdot W_V
\end{align}

where $W_Q$ $W_K$, and $W_V$ are trainable projection matrices. The Transformer encoder then effectively captures the global dependencies between features through the self-attention mechanism. The specific computation of the self-attention mechanism is as follows:
\begin{align}
	\text{Attention}(Q, K, V) = \text{softmax}\left( \frac{QK^T}{\sqrt{d_k}} \right) V
\end{align}

where $dk$ is the dimension of the key. The Transformer encoder processes the time-series features through multi-head attention, capturing the global information of the signal in the time domain, with an output dimension of $B \times 8 \times d\_{\textit{model}}$. This encoding of each time step's features captures the signal's global information in the time domain.

Finally, the model applies adaptive average pooling to reduce the dimensionality of the time-series features, resulting in a fixed-length time-series feature vector, which serves as the final output of the time-series branch.
\subsection{SCAPE}
The goal of the dynamic feature fusion module is to establish synergistic interactions between multimodal signals to effectively integrate information from different sources. To address the limitations of traditional feature fusion methods in spatial alignment and spatiotemporal differences, this paper proposes the SCAPE module, which is based on the DCAFE module introduced by Gupta et al. \cite{42}. The SCAPE module enhances real-time spatial encoding capabilities and incorporates a weighted calculation of channel attention to refine the attention weighting, thereby improving the model’s feature discriminative ability.

This attention mechanism, through the integration of positional encoders and spatial attention mechanisms, significantly enhances the expressive power of the signal features. The following is a detailed description of this module.
\subsubsection{\textbf{PAFF}}
First, the time-series and image data are concatenated along the channel dimension and then undergo a linear transformation, which allows for the maximal preservation of their respective features without increasing computational complexity. The processed mixed information is then fed into the initial part of the module, where the Position-Aware Feature Fusion (PAFF) is introduced. Through convolutional projection, spatial geometric information is directly fused into the feature representation.

The process starts by generating a relative position encoding tensor, which is projected using a 1×1 convolution, followed by standardization through BN and non-linear transformation using the ReLU activation function. Since the convolution operation is trainable, the position encoding injection process can be optimized during backpropagation, allowing the network to adjust the spatial encoding weights automatically. This enables flexible adjustment of the spatial feature fusion under different tasks or signal conditions, enhancing the model's generalization capability.

In mathematical terms, after introducing position encoding, the forward propagation process becomes:
\begin{align}
	f(x) = \sigma(W \cdot x + W \cdot P + b)
\end{align}

where $x$ represents the input feature vector, $W$ is the weight matrix, $b$ is the bias vector, $\sigma$ is the activation function, and $P$ denotes the position encoding. The gradient calculation is then transformed as:
\begin{align}
	\frac{\partial L}{\partial W} = \frac{\partial L}{\partial f} \cdot (x + P)
\end{align}

where $L$ is the loss function, and $f$ is the output of the convolutional layer.
By introducing the additional term $\frac{\partial L}{\partial f} \cdot P$, the model mitigates the issue of gradient vanishing, especially when the input feature values are small and gradients become weak. The position encoding expands the input value range, increasing the base magnitude of the gradient, thus ensuring effective weight updates. Additionally, the position encoding $P$ introduces spatial information into the gradient computation, enhancing the spatial distinguishability of the gradients. This makes differences more apparent when input features are similar. By guiding the model to learn the relationship between position and the task, the position encoding accelerates the understanding of spatial structures, thereby facilitating faster convergence and improving training speed. From the perspective of training efficiency, position encoding guides the model to learn the correlation between "spatial position and modulation recognition tasks," accelerating the understanding of the signal's spatial structure, reducing the number of feature exploration iterations, and achieving faster convergence while shortening the training period. The specific program implementation details are shown in Algorithm~\ref{algorithm}:
\begin{algorithm}[!htb]
\caption{PAFF Module}
\label{algorithm}
\begin{algorithmic}[1]
\State \textbf{Input:} Feature map $x \in \mathbb{R}^{n \times c \times h \times w}$
\State \textbf{Output:} Enhanced feature map with position encoding $f(x) \in \mathbb{R}^{n \times c \times h \times w}$

\State \textbf{Generate Position Encoding:}
\State $pos_{enc} \leftarrow \text{CreatePositionGrid}(h, w, n)$  \Comment{Create h×w position grid encoding}

\State \textbf{Process and Inject Position:}
\State $pos_{proj} \leftarrow \text{ReLU}(\text{BatchNorm}(\text{Conv}_{1\times1}(pos_{enc})))$
\State $f(x) \leftarrow x + pos_{proj}$

\State \textbf{Return} $f(x)$
\end{algorithmic}
\end{algorithm}
\subsubsection{\textbf{Coordinate Spatial Attention Mechanism}}
This module extends the input information from the plane into space, enhancing the ability to capture joint feature information through joint attention on both the spatial coordinates and the channels. Traditional coordinate attention mechanisms generally use average pooling, but since average pooling uses the mean feature vectors within the pooling window, it can cause the most important feature spaces to become blurred, leading to the aggregation of redundant or ambiguous information in the feature maps. In contrast, max pooling can capture important features. Additionally, in extremely low SNR environments, the introduction of channel attention enhances the non-linear interaction capability, forming a dual filtering mechanism that significantly improves the suppression of noise and irrelevant features.

Thus, this paper proposes a coordinate attention mechanism, combined with channel attention. The detailed process is described as follows: First, the feature vector with spatial encoding is input into a dual-branch structure. Let the dimensions of the feature vector be $H \times W \times C$ (where $H$ is the height, $W$ is the width, and $C$ is the number of channels). Then, attention SCAPE encodes the spatial context from both the horizontal and vertical directions. For the $c-th$ channel index, the horizontal and vertical spatial information of the coordinates are obtained as shown in the next formula.
\begin{align}
	f_a^c = \delta\left( \text{Conv}\left[ X\text{AvgPool}(F_h^c); Y\text{AvgPool}(F_w^c) \right] \right)
\end{align}
where,
\begin{align}
	X\text{AvgPool}(F_h^c) = \frac{1}{W} \sum_{i=0}^{W} F_c(h,i)
\end{align}
and     
\begin{align}
	Y\text{AvgPool}(F_w^c) = \frac{1}{H} \sum_{j=0}^{H} F_c(j, \omega) \quad \text{where } c = 1, 2, 3, \dots, C
\end{align}
\begin{align}
	\delta = \text{ReLU}(x)
\end{align}

Using $Conv(\cdot)$ to represent 1×1 convolution, $[;]$ denotes the splitting of features along the spatial dimension, resulting in two feature vectors of size $C \times H \times 1$ and $C \times 1 \times W$, respectively. XAvgPool and YAvgPool represent the average pooling operations performed along the height ($h$ direction) and width ($w$ direction) of the $c-th$ channel, capturing direction-sensitive information along the $X$ and $Y$ axes and integrating the contextual features of the spatial coordinates. The activation function used is ReLU, which, with its limited output range, enhances the model's generalization ability while significantly improving computational efficiency and numerical stability.

Next, the horizontally compressed vector $C \times H \times 1$ is reshaped and concatenated with $C \times 1 \times W$, resulting in an attention map $f_a^c \in C \times (H + W) \times 1$. Similarly, the attention map $f_m^c$ can also be obtained. This is followed by a 1×1 convolution operation, and then downsampled, reducing the number of channels to $C/Dr$. The specific formula is as follows:
\begin{align}
	C_{\text{out}} = \max\left( 8, C_{\text{in}} / D_r \right)
\end{align}

Next, the feature map $f_a \in \left( C / D_r \right) \times (H + W) \times 1$ is batch normalized and passed through ReLU. Then, it is split along the height dimension into $f_h^a \in \left( C / D_r \right) \times H \times 1$ and $f_w^a \in \left( C / D_r \right) \times W \times 1$. These two vectors are then processed through 1×1 convolutions and returned to the original channel count, so that $f_h^a \in C \times H \times 1$ and $f_w^a \in C \times W \times 1$. Finally, they are processed through a sigmoid function to obtain the attention weight matrix. The specific formula is as follows:
\begin{align}
	s_h^a = \sigma(\text{Conv}(f_h^a)) \quad \text{and} \quad s_w^a = \sigma(\text{Conv}(f_w^a))
\end{align}

In the expression, $s_h^a \in (C \times 1 \times H)$ and $s_w^a \in (C \times W \times 1)$ represent the attention weight matrices, and $\sigma$ denotes the Sigmoid activation function. Then, the channel attention matrix $s_c^a \in (C \times 1 \times 1)$ is obtained through the Sigmoid function.

Similarly, the SCAPE attention mechanism also uses max pooling to perform the same operation, resulting in the attention weight matrices $s_h^m$ and $s_w^m$, which capture important features. The original feature vector is then processed through global average pooling to obtain a tensor of size $C \times 1 \times 1$. It is then matched in shape through a 1×1 convolution, and the final channel attention matrix $s_c^m \in (C \times 1 \times 1)$ is obtained using the Sigmoid function.

Finally, the attention weight matrix is multiplied by the original feature vector to obtain the final feature vector ($Y$), as shown in the formula.
\begin{align}
	Y^m(i, j, k) = F(i, j, k) \odot s_h^m(i) \odot s_w^m(j) \odot s_c^m(k)
\end{align}
\begin{align}
	Y^a(i, j, k) = F(i, j, k) \odot s_h^a(i) \odot s_w^a(j) \odot s_c^a(k)
\end{align}
\begin{align}
	Y(i, j) = \text{Concat}(Y^m(i, j, k), Y^a(i, j, k))
\end{align}

In this module, the computed channel weights are applied through a per-channel multiplication operation with the dual-branch aggregated features, achieving a channel-level recombination. This recombination mechanism adaptively adjusts the contribution of each channel, ultimately producing enhanced output feature maps. Especially in complex, low SNR communication environments, the channel attention mechanism automatically learns which channels to prioritize and the optimal fusion weights, ensuring that the features from different modalities are properly weighted and combined, thereby improving overall classification performance.
\section{Experiment} \label{Evaluation Dataset}
\subsection{Evaluation Dataset}
\begin{table*}[htbp]
  \centering
  \caption{Dataset Summary for Modulation Classification}
  \small 
  \begin{tabular}{l>{\raggedright\arraybackslash}p{3.5cm}>{\raggedright\arraybackslash}p{3.5cm}>{\raggedright\arraybackslash}p{3.5cm}}
    \toprule
    Dataset & RadioML2016.10a & RadioML2016.10b & HisarMod2019.1 \\
    \midrule
    Modulation Schemes & 11 classes (8PSK, BPSK, CPFSK, GFSK, PAM4, 16QAM, AM-DSB, 64QAM, QPSK, WBFM) & 10 classes (8PSK, BPSK, CPFSK, GFSK, PAM4, 16QAM, 64QAM, QPSK, WBFM) & 26 classes (AM-DSB, AM-USB, AM-LSB, FM, 2FSK, 4FSK, 8FSK, 16FSK, 4PAM, 8PAM, 16PAM, BPSK, QPSK, 8PSK, 16PSK, 32PSK, 64PSK, 4QAM, 8QAM, 16QAM, 32QAM, 64QAM, 256QAM, 128QAM) \\
    \midrule
    Sample Dimension & 2×128 & 2×128 & 2×1024 \\
    \midrule
    Dataset Size & 220000 & 1200000 & 780000 \\
    \midrule
    SNR Range(dB) & -20:2:18 & -20:2:18 & -20:2:18 \\
    \midrule
    Channel model & center frequency offset (CFO), sample rate offset (SRO), additive white Gaussian noise(AWGN), Rayleigh, Rician & CFO, SRO, AWGN, Rayleigh, Rician & five types of channels: ideal, static, Rayleigh, Rician (k=3), and Nakagami-m (m=2) \\
    \midrule
    Data source & GNU Radio simulated & GNU Radio simulated & MATLAB 2017a simulated \\
    \bottomrule
  \end{tabular}
  \label{tab:dataset_summary}
\end{table*}
This section systematically and comprehensively evaluates the proposed AMR method using publicly available datasets: RadioML2016.10a \cite{18}, RadioML2016.10b, and HisarMod2019.1 \cite{43}. The related parameters are illustrated in Table~\ref{tab:dataset_summary}. These datasets, generated via GNU Radio and MATLAB 2017a, are designed to simulate complex and challenging wireless transmission environments, covering various channel distortions and interference factors, thus validating the performance of the AMR method in real-world communication scenarios.

Specifically, the RadioML2016.10a dataset includes 11 different modulation schemes, namely: 8-phase shift keying (8PSK), binary phase shift keying (BPSK), 16-quadrature amplitude modulation (QAM16), 64-quadrature amplitude modulation (QAM64), quadrature phase shift keying (QPSK), wideband frequency modulation (WBFM), continuous phase frequency shift keying (CPFSK), Gaussian frequency shift keying (GFSK), amplitude modulation double sideband (AM-DSB), amplitude modulation single sideband (AM-SSB), and 4-level pulse amplitude modulation (PAM4). The SNR in this dataset ranges from -20 dB to 18 dB with a step size of 2 dB. Each modulation scheme contains 1000 samples at each SNR level, with each sample comprising I and Q signals. Noise and frequency offset distortions are introduced in the data, with a sampling rate offset standard deviation of 0.01 Hz, a maximum offset of 50 Hz, and a maximum carrier frequency offset of 500 Hz.

In contrast, the RadioML2016.10b dataset is an extended version of RadioML2016.10a, which removes the AM-SSB modulation scheme and significantly increases the sample size for each modulation scheme at each SNR level to 6000 samples. Additionally, each sample in the RadioML2016.10b dataset is formatted as 2×128, meaning each sample consists of two channels, each containing 128 data points, thus enhancing the scale and diversity of the dataset.

Furthermore, the HisarMod2019.1 dataset, generated with MATLAB 2017a, contains signals transmitted under five different channel conditions: static, Rayleigh, ideal, Nakagami-m (m=2), and Rician (k=3). This dataset includes 780,000 samples and covers 26 modulation schemes, with each sample formatted as 2×1024 (i.e., two channels, each containing 1024 data points). These design choices enable HisarMod2019.1 to more accurately simulate wireless transmission characteristics under various channel environments, making it suitable for validating the robustness and generalization capabilities of AMR methods under diverse channel conditions.

For the benchmark datasets RadioML2016.10a, RadioML2016.10b, and HisarMod2019.1, the datasets were randomly split into training, validation, and testing sets with a 7:2:1 ratio to ensure the representativeness and reliability of the experimental results. All experiments were conducted on a Linux platform equipped with a GeForce GTX 4090 GPU to fully leverage the hardware's computational capabilities, ensuring both efficiency and stability during the experiments.

The experimental hyperparameter configuration is as follows: the initial learning rate is set to 0.00005, and L2 regularization is introduced to control model complexity and prevent overfitting. The weight decay coefficient is set to 0.01, which helps apply appropriate constraints to the parameters during optimization, reducing overfitting to the training data. The model optimization utilizes the AdamW optimizer, which, based on Adam, explicitly controls weight decay. This mechanism effectively mitigates overfitting in high-dimensional parameter spaces, particularly suitable for multimodal learning tasks.

For the loss function, the MCANet model adopts the cross-entropy loss, which is the standard choice for multi-class problems. The cross-entropy loss assigns corresponding probabilities to each class in a multi-class task, facilitating effective optimization and ensuring consistency between the target and the optimization direction during training.

Through in-depth analysis and evaluation of these three datasets, along with a well-designed hyperparameter configuration, these datasets provide the correct direction for model training and leverages a rich set of datasets to validate its performance.

\subsection{Comparison with Mainstream Deep Learning Models}
\begin{table*}[tb]
	\centering
	\caption{COMPARISON OF MODEL PERFORMANCE ON THREE DATASETS}
	\label{Comparison with Mainstream Deep Learning Models}
	\small
	\begin{tabular}{cccc}
		\toprule
		Model & Dataset & Highest accuracy & Overall accuracy \\ 
		\midrule
		PET-CGDNN & RadioML2016.10a & 90.63\% & 60.12\% \\
		   & RadioML2016.10b & 93.39\% & 63.87\% \\
		   & HisarMod2019.1 & 89.33\% & 67.21\% \\ 
		\midrule
		AMC-NET & RadioML2016.10a & 92.78\% & 62.29\% \\
		   & RadioML2016.10b & 93.81\% & 65.09\% \\
    		 & HisarMod2019.1 & 99.40\% & 72.67\% \\ 
		\midrule
		FEA-T & RadioML2016.10a & 90.06\% & 60.44\% \\
		   & RadioML2016.10b & 93.78\% & 64.31\% \\
		   & HisarMod2019.1 & 99.86\% & 70.21\% \\ 
		\midrule
		MCLDNN & RadioML2016.10a & 92.75\% & 61.82\% \\
		   & RadioML2016.10b & 93.77\% & 64.43\% \\
		   & HisarMod2019.1 & 97.95\% & 68.53\% \\ 
		\midrule
		MCANet (Ours) & RadioML2016.10a & \textbf{97.57\%} & \textbf{66.12\%} \\
		   & RadioML2016.10b & \textbf{98.89\%} & \textbf{66.53\%} \\
		   & HisarMod2019.1 & \textbf{99.96\%} & \textbf{76.58\%} \\
		\bottomrule
	\end{tabular}
\end{table*}

This section compares the proposed MCANet model with several existing DL-AMR models, including MCLDNN \cite{44}, PET-CGDNN \cite{45}, AMC-NET \cite{46}, and FEA-T \cite{47}. The results are shown in Table~\ref{Comparison with Mainstream Deep Learning Models}.
\raggedbottom

On the RadioML2016.10a dataset, the MCANet model achieved outstanding performance, with a peak accuracy of 97.57\%. This result demonstrates the superior capability of the model and significantly outperforms mainstream algorithms. Compared with PET-CGDNN (90.63\%), AMC-NET (92.78\%), and FEA-T (90.06\%), the MCANet model shows a clear advantage in classification accuracy. Moreover, it also achieved an overall accuracy of 66.12\%, representing a substantial improvement over PET-CGDNN (60.12\%) and MCLDNN (61.82\%), further validating the effectiveness and competitiveness of the MCANet model on this dataset.
\raggedbottom

On the RadioML2016.10b dataset, the MCANet model exhibited similar performance, achieving a peak accuracy of 98.89\%. In comparison, although PET-CGDNN (93.39\%) and AMC-NET (93.81\%) were previously regarded as strong performers, their results on this dataset remain slightly inferior to those of the MCANet model. In terms of overall accuracy, the MCANet model achieved 66.53\%, clearly outperforming PET-CGDNN and other baseline models.
\raggedbottom

The experimental results on the HisarMod2019.1 dataset further verified the superiority of the MCANet model. On this dataset, the MCANet model achieved an impressive 99.96\% peak accuracy, outperforming other models, particularly AMC-NET (99.86\%) and PET-CGDNN (89.33\%). In terms of overall accuracy, the MCANet model also demonstrated strong performance, reaching 76.58\%, which substantially exceeds PET-CGDNN (67.21\%) and MCLDNN (68.53\%). This result confirms the generalization capability of the model across different datasets.

\begin{figure}[H]
    \centering
    \includegraphics[width=\columnwidth]{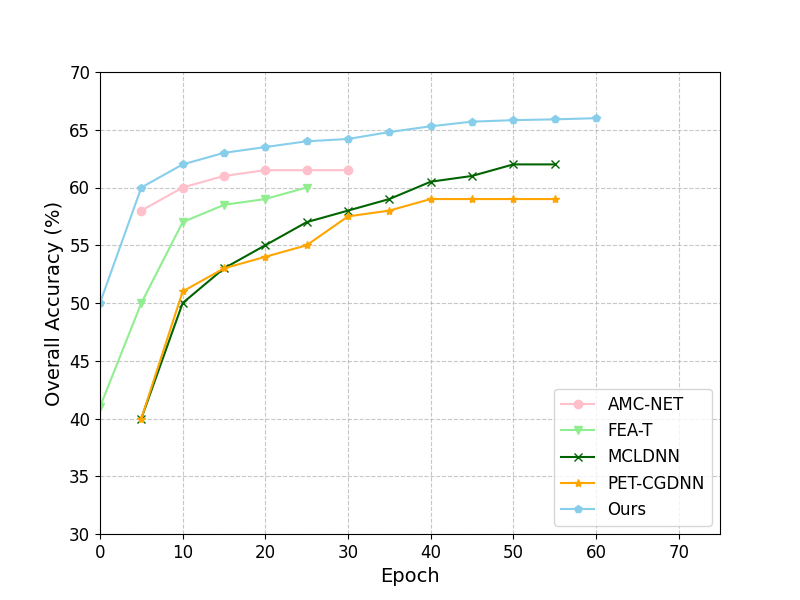}
    \caption{Epoch vs. Accuracy Comparison: MCANet vs. Mainstream Models} 
    \label{Epoch vs. Accuracy Comparison: MCANet vs. Mainstream Models}
\end{figure}

In terms of convergence speed, the results, as shown in Fig.~\ref{Epoch vs. Accuracy Comparison: MCANet vs. Mainstream Models}, indicate that the MCANet model exhibits a relatively fast convergence in the early stages of training on the RadioML2016.10a dataset. However, as the number of training epochs increases, the convergence rate gradually slows down. Compared to other mainstream models, the MCANet model requires an additional 15 to 20 epochs to achieve optimal recognition accuracy.
Overall, despite the MCANet model’s slower convergence speed, it outperforms current state-of-the-art deep learning models in both peak and overall accuracy, demonstrating its significant advantages and broad application potential in modulation classification tasks.

\subsection{Modulation analysis}
\begin{figure*}[tb!]
    \centering
    \includegraphics[width=\textwidth]{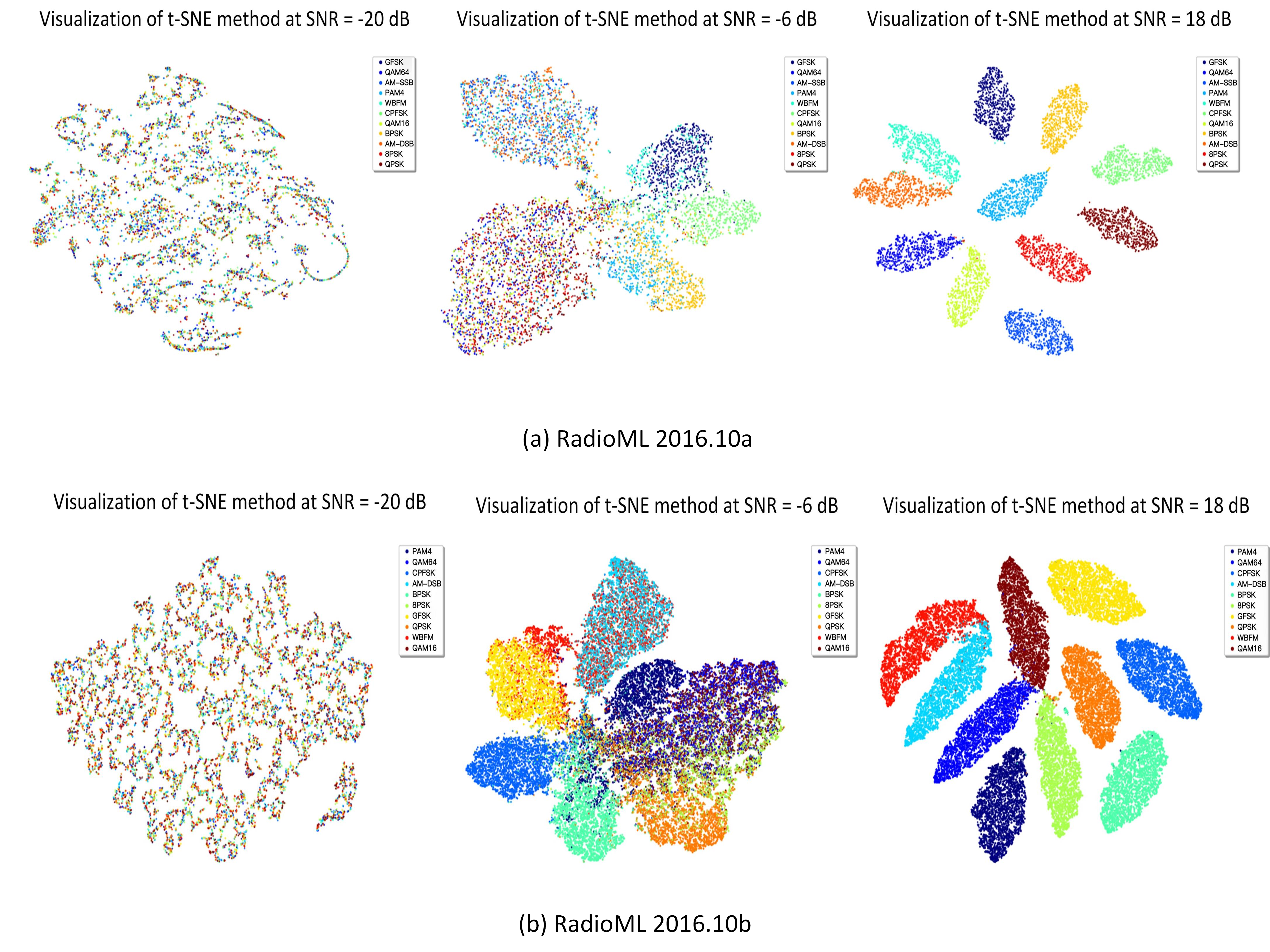}
    \caption{t-SNE Visualization of Feature Space Distribution for Modulation Classification} 
    \label{t-SNE Visualization of Feature Space Distribution for Modulation Classification}
\end{figure*}

This section elaborates on the experimental design and results of the proposed modulation classification method, covering the overview of evaluation datasets and the verification of the method’s effectiveness across different datasets. As shown in the Fig.~\ref{t-SNE Visualization of Feature Space Distribution for Modulation Classification}, on the RadioML2016.10a dataset, when the SNR is -20 dB, the features extracted by the MCANet model are almost entirely overwhelmed by noise, resulting in highly dispersed samples in the feature space. This indicates that, under low SNR conditions, noise significantly interferes with the model’s feature learning, making it difficult for the model to effectively recognize and distinguish different modulation formats on the RadioML2016.10a dataset.
\begin{figure*}[tb!]
    \centering
    \includegraphics[width=\textwidth]{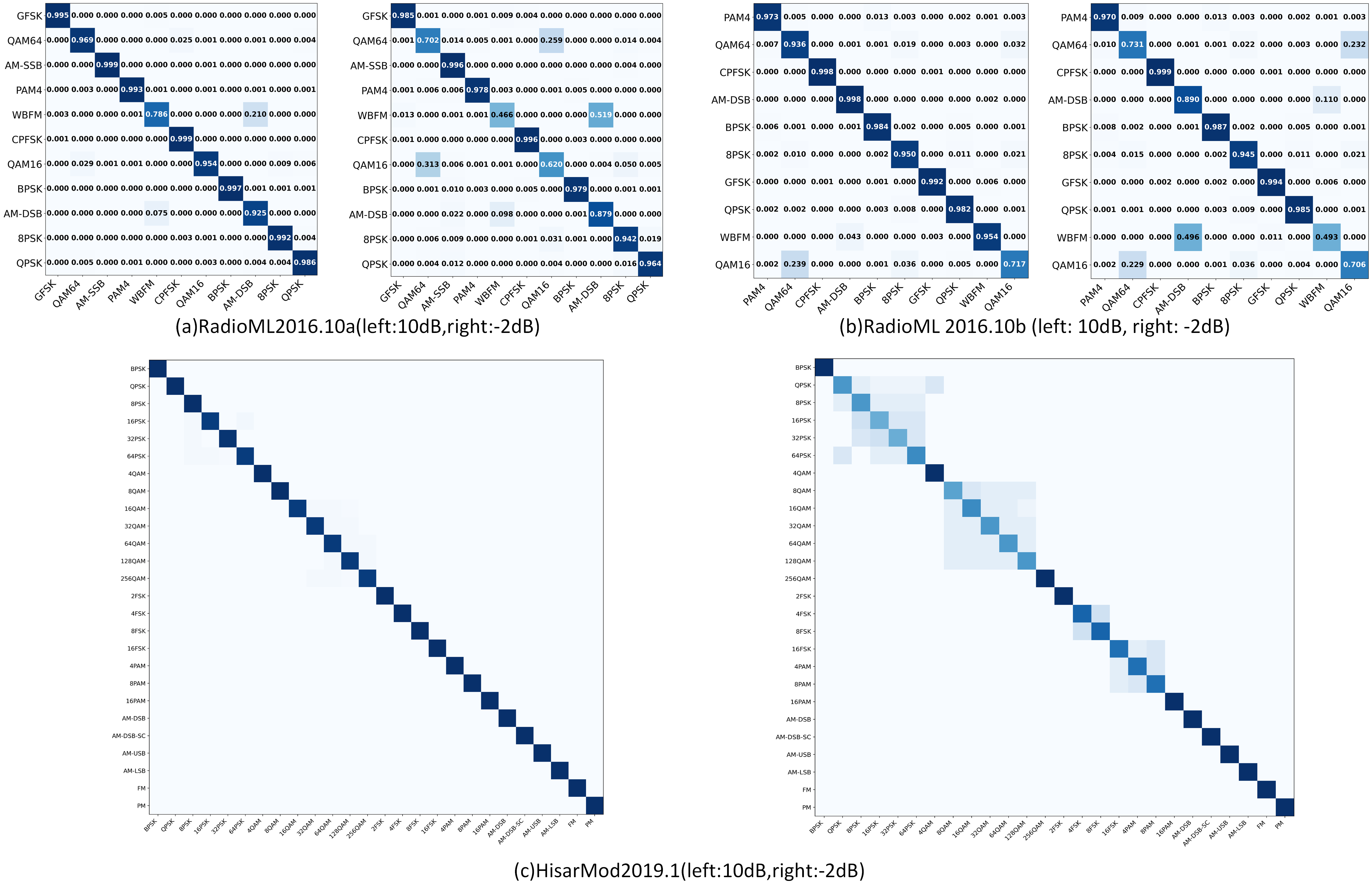}
    \caption{Normalized Confusion Matrix for RML2016 Series and HisarMod2019.1 Datasets} 
    \label{Normalized Confusion Matrix for RML2016 Series and HisarMod2019.1 Datasets}
\end{figure*}
As the SNR increases to -6 dB, features from modulation schemes such as GFSK and CPFSK begin to form relatively independent clusters in the lower-dimensional space after dimensionality reduction on the RadioML2016.10a dataset, showing improved separability. At this point, the MCANet model is able to extract effective features from the noise, enhancing the distinguishability between categories. Although there is still some overlap between the clusters of WBFM and AM-DSB, as well as between PAM4 and BPSK, indicating higher similarity in their feature space distributions, this does not hinder their ability to maintain a certain level of separation under low SNR conditions.

When the SNR is further increased to 18 dB, the features of most categories in the RadioML2016.10a dataset form clear boundaries in the feature space, significantly enhancing the separability between categories. Notably, although there remains some intra-class confusion between QAM16 and QAM64, this confusion is significantly alleviated as the SNR increases to 18 dB, and the separation between the two modulation schemes in the feature space is notably improved.

Next, we analyze the RadioML2016.10b dataset to evaluate the model's performance under different SNR conditions. On the RadioML2016.10b dataset, when the SNR is -20 dB, the classification performance of the MCANet model approaches random guessing. As the SNR increases to -6 dB, features from modulation schemes such as CPFSK and 8PSK begin to form relatively independent clusters in the lower-dimensional space after dimensionality reduction. However, the feature clusters of WBFM and AM-DSB show some overlap in the reduced space, indicating that their distributions in feature space are not entirely independent. When the SNR is further increased to 18 dB, the features of most categories in the RadioML2016.10b dataset form clear boundaries in the feature space, demonstrating the model's strong feature extraction capability.

Fig.~\ref{Normalized Confusion Matrix for RML2016 Series and HisarMod2019.1 Datasets} illustrates the classification performance of the MCANet model on different modulation formats in the RML2016 series and HisarMod2019.1 datasets. At an SNR of -2 dB, significant interference from the signal model still exists, including amplitude and phase random perturbations caused by AWGN, as well as phase noise and co-channel interference from adjacent channels. These interferences collectively cause severe distortion of the signal waveforms in the RML2016 series dataset, leading to confusion between WBFM and AM-DSB due to the blurring of their envelope features, resulting in misclassification of these two modulation formats. Moreover, the noise overwhelms the constellation point features, making it difficult to distinguish between them. As a result, the model also exhibits some degree of confusion between QAM16 and QAM64. When the SNR is increased to 10 dB, almost all of the confusion is eliminated. However, a certain degree of misclassification still remains between QAM16 and QAM64, indicating that the model's ability to distinguish between high-order modulation formats with similar characteristics is still limited.

Additionally, Fig.~\ref{Normalized Confusion Matrix for RML2016 Series and HisarMod2019.1 Datasets}(c) presents the confusion matrix results for the HisarMod2019.1 dataset. At an SNR of -2 dB, misclassifications primarily occur within modulation type families. When the SNR is raised to 10 dB, the misclassifications nearly disappear, further validating the MCANet model's classification capability under low SNR conditions.

\subsection{Ablation Study}
\begin{table}[H]
  \centering
  \caption{THE ABLATION RESULTS OF MCANet}
  \label{THE ABLATION RESULTS OF MCANet}
  \begin{tabular}{llc}
    \toprule 
    Datasets & Model & Overall \\
    \midrule 
    RadioML2016.10a & MCANet (Ours) & 66.12\% \\
    & w/o PAFF& 66.07\% \\
    & w/o WaveFilter & 64.73\% \\
    & w/o Fusion Gate & 65.57\% \\
    RadioML2016.10b & MCANet (Ours) & 66.53\% \\
    & w/o PAFF& 66.21\% \\
    & w/o WaveFilter & 65.16\% \\
    & w/o Fusion Gate & 65.87\% \\
    \bottomrule 
  \end{tabular}
\end{table}
\begin{figure}[H]
	\centering
	\includegraphics[width=\columnwidth]{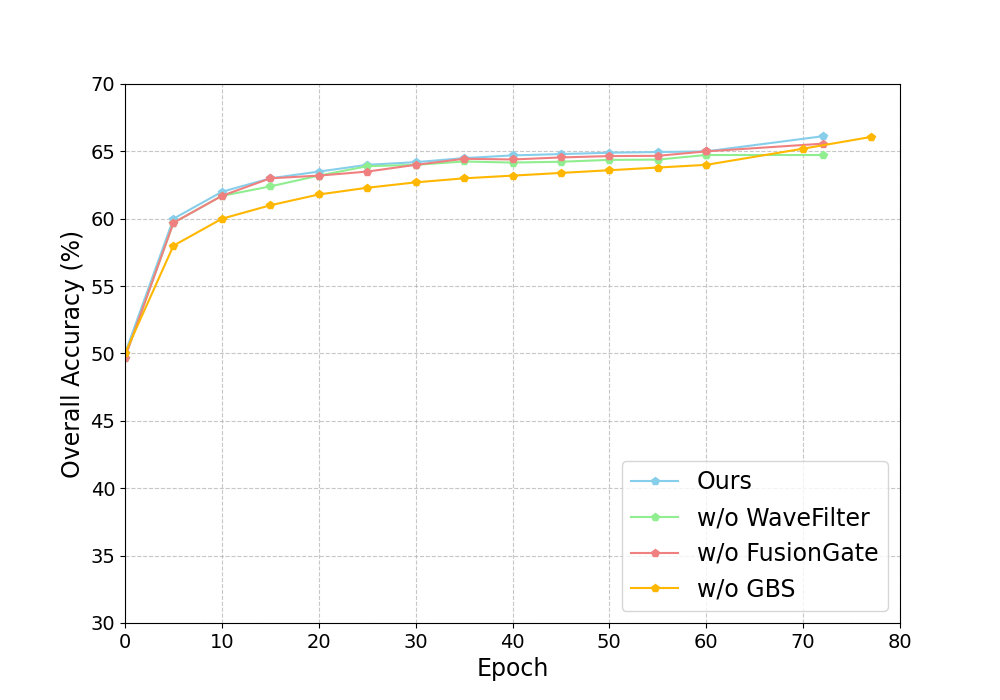}
	\caption{Ablation Impact on Accuracy vs. Epochs on the RadioML2016.10a dataset} 
	\label{Ablation Impact on Accuracy vs. Epochs on the RadioML2016.10a dataset}
\end{figure}
This section evaluates the contribution of each module to the model’s performance by conducting ablation experiments that validate the enhancement provided by the core modules. Ablation experiments were performed on two datasets by removing the PAFF module, the WaveFilter module for learnable high-frequency noise suppression, and the FusionGate module for linear weight control of constellation and eye diagrams.
\raggedbottom

The results, as shown in Table~\ref{THE ABLATION RESULTS OF MCANet}, reveal similar trends on both the RML2016.10a and RML2016.10b datasets. The removal of the WaveFilter and FusionGate modules caused varying degrees of accuracy degradation, indicating that their removal impaired the model’s feature extraction capability. 
\raggedbottom

Although the PAFF module contributed little to accuracy, as shown in Fig.~\ref{Ablation Impact on Accuracy vs. Epochs on the RadioML2016.10a dataset}, it significantly accelerated the model's convergence speed and improved training efficiency. It is evident that the absence of these three components weakens the model's feature processing ability, thereby confirming their positive impact on model performance.
\section{Conclusion} \label{sec:Conclusion}
This paper proposes a novel AMR framework based on multimodal deep learning. The model synergistically integrates signal features from the time domain, frequency domain, and image modalities, overcoming the performance bottleneck of traditional unimodal methods, and significantly improving classification accuracy and robustness in low SNR and complex channel environments. The study shows that multimodal learning effectively extracts complementary representations of the signal in different domains, and through multi-modal semantic alignment and adaptive feature integration, it provides a more powerful solution for signal recognition in complex electromagnetic environments. However, the model does have some disadvantages in terms of computational complexity. Nevertheless, with advancements in hardware technology, modern high-performance devices, especially GPUs and dedicated accelerators, can effectively support such complex models. Therefore, despite the model's large size, current computational capabilities are sufficient to meet its training and inference requirements, further validating the tremendous potential and prospects of multimodal fusion in automatic modulation recognition tasks.

Based on the results of this study, we will further explore lightweight multimodal network design, aiming to reduce computational overhead while maintaining accuracy, and extend the application of multimodal fusion in dynamic spectrum access, intelligent interference mitigation, and other communication scenarios.

\bibliographystyle{IEEEtran}
\bibliography{refs2}

\begin{thebibliography}{10}
\providecommand{\url}[1]{#1}
\csname url@samestyle\endcsname
\providecommand{\newblock}{\relax}
\providecommand{\bibinfo}[2]{#2}
\providecommand{\BIBentrySTDinterwordspacing}{\spaceskip=0pt\relax}
\providecommand{\BIBentryALTinterwordstretchfactor}{4}
\providecommand{\BIBentryALTinterwordspacing}{\spaceskip=\fontdimen2\font plus
\BIBentryALTinterwordstretchfactor\fontdimen3\font minus
  \fontdimen4\font\relax}
\providecommand{\BIBforeignlanguage}[2]{{%
\expandafter\ifx\csname l@#1\endcsname\relax
\typeout{** WARNING: IEEEtran.bst: No hyphenation pattern has been}%
\typeout{** loaded for the language `#1'. Using the pattern for}%
\typeout{** the default language instead.}%
\else
\language=\csname l@#1\endcsname
\fi
#2}}
\providecommand{\BIBdecl}{\relax}
\BIBdecl

\bibitem{1}
C.~Liu, Z.~Cai, and B.~Zhang, ``Automatic modulation classification based on
  complex-valued convolutional neural network and semi-supervised learning,''
  in \emph{2023 IEEE 23rd International Conference on Communication Technology
  (ICCT)}.\hskip 1em plus 0.5em minus 0.4em\relax IEEE, 2023, pp. 241--246.

\bibitem{2}
J.~Jiang, Z.~Wang, H.~Zhao, S.~Qiu, and J.~Li, ``Modulation recognition method
  of satellite communication based on cldnn model,'' in \emph{2021 IEEE 30th
  International Symposium on Industrial Electronics (ISIE)}.\hskip 1em plus
  0.5em minus 0.4em\relax IEEE, 2021, pp. 1--6.

\bibitem{3}
Y.~A. Eldemerdash, O.~A. Dobre, and M.~{\"O}ner, ``Signal identification for
  multiple-antenna wireless systems: Achievements and challenges,'' \emph{IEEE
  Communications Surveys \& Tutorials}, vol.~18, no.~3, pp. 1524--1551, 2016.

\bibitem{4}
J.~L. Xu, W.~Su, and M.~Zhou, ``Likelihood-ratio approaches to automatic
  modulation classification,'' \emph{IEEE Transactions on Systems, Man, and
  Cybernetics, Part C (Applications and Reviews)}, vol.~41, no.~4, pp.
  455--469, 2010.

\bibitem{5}
S.~Huang, Y.~Yao, Z.~Wei, Z.~Feng, and P.~Zhang, ``Automatic modulation
  classification of overlapped sources using multiple cumulants,'' \emph{IEEE
  Transactions on Vehicular Technology}, vol.~66, no.~7, pp. 6089--6101, 2016.

\bibitem{6}
V.~Iglesias, J.~Grajal, P.~Royer, M.~A. Sanchez, M.~Lopez-Vallejo, and O.~A.
  Yeste-Ojeda, ``Real-time low-complexity automatic modulation classifier for
  pulsed radar signals,'' \emph{IEEE Transactions on Aerospace and Electronic
  Systems}, vol.~51, no.~1, pp. 108--126, 2015.

\bibitem{7}
S.~Majhi, R.~Gupta, W.~Xiang, and S.~Glisic, ``Hierarchical hypothesis and
  feature-based blind modulation classification for linearly modulated
  signals,'' \emph{IEEE Transactions on Vehicular Technology}, vol.~66, no.~12,
  pp. 11\,057--11\,069, 2017.

\bibitem{8}
D.~Zibar, O.~Winther, N.~Franceschi, R.~Borkowski, A.~Caballero, V.~Arlunno,
  M.~N. Schmidt, N.~G. Gonzales, B.~Mao, Y.~Ye \emph{et~al.}, ``Nonlinear
  impairment compensation using expectation maximization for dispersion managed
  and unmanaged pdm 16-qam transmission,'' \emph{Optics express}, vol.~20,
  no.~26, pp. B181--B196, 2012.

\bibitem{9}
C.~L. KUMAR, ``Modulation classification in fading channels using expectation
  maximization,'' Ph.D. dissertation, INDIAN INSTITUTE OF TECHNOLOGY MADRAS,
  2021.

\bibitem{10}
D.~Wang, M.~Zhang, Z.~Li, Y.~Cui, J.~Liu, Y.~Yang, and H.~Wang, ``Nonlinear
  decision boundary created by a machine learning-based classifier to mitigate
  nonlinear phase noise,'' in \emph{2015 European Conference on Optical
  Communication (ECOC)}.\hskip 1em plus 0.5em minus 0.4em\relax IEEE, 2015, pp.
  1--3.

\bibitem{11}
D.~Wang, M.~Zhang, Z.~Cai, Y.~Cui, Z.~Li, H.~Han, M.~Fu, and B.~Luo,
  ``Combatting nonlinear phase noise in coherent optical systems with an
  optimized decision processor based on machine learning,'' \emph{Optics
  Communications}, vol. 369, pp. 199--208, 2016.

\bibitem{12}
X.~Zhang, J.~Sun, and X.~Zhang, ``Automatic modulation classification based on
  novel feature extraction algorithms,'' \emph{IEEE Access}, vol.~8, pp.
  16\,362--16\,371, 2020.

\bibitem{13}
L.~Barletta, A.~Giusti, C.~Rottondi, and M.~Tornatore, ``Qot estimation for
  unestablished lighpaths using machine learning,'' in \emph{Optical Fiber
  Communication Conference}.\hskip 1em plus 0.5em minus 0.4em\relax Optica
  Publishing Group, 2017, pp. Th1J--1.

\bibitem{14}
A.~Ali, T.~M. Hasan, and S.~D. Mohammed, ``Digital modulation classification
  based on chicken swarm optimization and random forest,'' \emph{J. Eng. Sci.
  Technol}, vol.~17, pp. 2095--2103, 2022.

\bibitem{15}
X.~Sun, S.~Su, Z.~Zuo, X.~Guo, and X.~Tan, ``Modulation classification using
  compressed sensing and decision tree--support vector machine in cognitive
  radio system,'' \emph{Sensors}, vol.~20, no.~5, p. 1438, 2020.

\bibitem{16}
Y.-Y. Song and Y.~Lu, ``Decision tree methods: applications for classification
  and prediction,'' \emph{Shanghai archives of psychiatry}, 2015.

\bibitem{17}
A.~K. Ali and E.~Ercelebi, ``Algorithm for automatic recognition of psk and qam
  with unique classifier based on features and threshold levels,'' \emph{ISA
  transactions}, vol. 102, pp. 173--192, 2020.

\bibitem{18}
T.~J. O'shea and N.~West, ``Radio machine learning dataset generation with gnu
  radio,'' in \emph{Proceedings of the GNU radio conference}, vol.~1, no.~1,
  2016.

\bibitem{19}
A.~Krizhevsky, I.~Sutskever, and G.~E. Hinton, ``Imagenet classification with
  deep convolutional neural networks,'' \emph{Advances in neural information
  processing systems}, vol.~25, 2012.

\bibitem{20}
F.~Zhao, ``Comparative analysis of cnn and resnet for automatic modulation
  recognition across various snr environments,'' \emph{Procedia Computer
  Science}, vol. 266, pp. 1433--1441, 2025.

\bibitem{21}
J.~Zhao, Q.~Cheng, H.~Wang, and Y.-D. Yao, ``Vit-mae based foundation model for
  automatic modulation classification,'' in \emph{2024 33rd Wireless and
  Optical Communications Conference (WOCC)}.\hskip 1em plus 0.5em minus
  0.4em\relax IEEE, 2024, pp. 50--54.

\bibitem{22}
A.~Faysal, M.~Rostami, R.~G. Roshan, H.~Wang, and N.~Muralidhar, ``Nmformer: A
  transformer for noisy modulation classification in wireless communication,''
  in \emph{2024 33rd Wireless and Optical Communications Conference
  (WOCC)}.\hskip 1em plus 0.5em minus 0.4em\relax IEEE, 2024, pp. 103--108.

\bibitem{23}
D.~Hong, Z.~Zhang, and X.~Xu, ``Automatic modulation classification using
  recurrent neural networks,'' in \emph{2017 3rd IEEE international conference
  on computer and communications (ICCC)}.\hskip 1em plus 0.5em minus
  0.4em\relax IEEE, 2017, pp. 695--700.

\bibitem{24}
S.~Rajendran, W.~Meert, D.~Giustiniano, V.~Lenders, and S.~Pollin, ``Deep
  learning models for wireless signal classification with distributed low-cost
  spectrum sensors,'' \emph{IEEE Transactions on Cognitive Communications and
  Networking}, vol.~4, no.~3, pp. 433--445, 2018.

\bibitem{25}
Z.~Ke and H.~Vikalo, ``Real-time radio technology and modulation classification
  via an lstm auto-encoder,'' \emph{IEEE Transactions on Wireless
  Communications}, vol.~21, no.~1, pp. 370--382, 2021.

\bibitem{26}
N.~E. West and T.~O'shea, ``Deep architectures for modulation recognition,'' in
  \emph{2017 IEEE international symposium on dynamic spectrum access networks
  (DySPAN)}.\hskip 1em plus 0.5em minus 0.4em\relax IEEE, 2017, pp. 1--6.

\bibitem{27}
D.~Luo and X.~Wang, ``Deformabletst: Transformer for time series forecasting
  without over-reliance on patching,'' \emph{Advances in Neural Information
  Processing Systems}, vol.~37, pp. 88\,003--88\,044, 2024.

\bibitem{28}
A.~Shabani, A.~Abdi, L.~Meng, and T.~Sylvain, ``Scaleformer: Iterative
  multi-scale refining transformers for time series forecasting,'' \emph{arXiv
  preprint arXiv:2206.04038}, 2022.

\bibitem{29}
R.~Ilbert, A.~Odonnat, V.~Feofanov, A.~Virmaux, G.~Paolo, T.~Palpanas, and
  I.~Redko, ``Unlocking the potential of transformers in time series
  forecasting with sharpness-aware minimization and channel-wise attention,''
  \emph{CoRR}, 2024.

\bibitem{30}
Z.~Liu, Y.~Lin, Y.~Cao, H.~Hu, Y.~Wei, Z.~Zhang, S.~Lin, and B.~Guo, ``Swin
  transformer: Hierarchical vision transformer using shifted windows,'' in
  \emph{Proceedings of the IEEE/CVF international conference on computer
  vision}, 2021, pp. 10\,012--10\,022.

\bibitem{31}
Z.~Chen, L.~Xie, J.~Niu, X.~Liu, L.~Wei, and Q.~Tian, ``Visformer: The
  vision-friendly transformer,'' in \emph{Proceedings of the IEEE/CVF
  international conference on computer vision}, 2021, pp. 589--598.

\bibitem{32}
R.~Zeng, X.~Chen, Y.~Pu, X.~Zhang, T.~Du, and S.~Ji, ``Clibe: detecting dynamic
  backdoors in transformer-based nlp models,'' \emph{arXiv preprint
  arXiv:2409.01193}, 2024.

\bibitem{33}
J.~Alammar, ``Ecco: An open source library for the explainability of
  transformer language models,'' in \emph{Proceedings of the 59th annual
  meeting of the association for computational linguistics and the 11th
  international joint conference on natural language processing: System
  demonstrations}, 2021, pp. 249--257.

\bibitem{34}
J.~Cai, F.~Gan, X.~Cao, and W.~Liu, ``Signal modulation classification based on
  the transformer network,'' \emph{IEEE Transactions on Cognitive
  Communications and Networking}, vol.~8, no.~3, pp. 1348--1357, 2022.

\bibitem{35}
W.~Kong, X.~Jiao, Y.~Xu, B.~Zhang, and Q.~Yang, ``A transformer-based
  contrastive semi-supervised learning framework for automatic modulation
  recognition,'' \emph{IEEE Transactions on Cognitive Communications and
  Networking}, vol.~9, no.~4, pp. 950--962, 2023.

\bibitem{36}
Y.~Tu, Y.~Lin, C.~Hou, and S.~Mao, ``Complex-valued networks for automatic
  modulation classification,'' \emph{IEEE Transactions on Vehicular
  Technology}, vol.~69, no.~9, pp. 10\,085--10\,089, 2020.

\bibitem{37}
A.~Samarkandi, A.~Almarhabi, H.~Alhazmi, and Y.-D. Yao, ``Combined signal
  representations for modulation classification using deep learning: Ambiguity
  function, constellation diagram, and eye diagram,'' in \emph{2023 32nd
  Wireless and Optical Communications Conference (WOCC)}.\hskip 1em plus 0.5em
  minus 0.4em\relax IEEE, 2023, pp. 1--4.

\bibitem{38}
W.~Zhao and Q.~Luo, ``Leveraging computer vision for automatic modulation
  classification: Insights from spectrum and constellation diagram analysis,''
  in \emph{International Conference on Pattern Recognition}.\hskip 1em plus
  0.5em minus 0.4em\relax Springer, 2024, pp. 284--294.

\bibitem{39}
L.~Zhuang, K.~Luo, and Z.~Yang, ``A multimodal gated recurrent unit neural
  network model for damage assessment in cfrp composites based on lamb waves
  and minimal sensing,'' \emph{IEEE Transactions on Instrumentation and
  Measurement}, vol.~73, pp. 1--11, 2024.

\bibitem{40}
J.~Han, Z.~Yu, and J.~Yang, ``Multimodal attention-based deep learning for
  automatic modulation classification,'' \emph{Frontiers in Energy Research},
  vol.~10, p. 1041862, 2022.

\bibitem{41}
K.~He, X.~Zhang, S.~Ren, and J.~Sun, ``Deep residual learning for image
  recognition,'' in \emph{Proceedings of the IEEE conference on computer vision
  and pattern recognition}, 2016, pp. 770--778.

\bibitem{42}
S.~Gupta and A.~K. Tripathi, ``Flora-net: Integrating dual coordinate attention
  with adaptive kernel based convolution network for medicinal flower
  identification,'' \emph{Computers and Electronics in Agriculture}, vol. 230,
  p. 109834, 2025.

\bibitem{43}
K.~Tekb{\i}y{\i}k, A.~R. Ekti, A.~G{\"o}r{\c{c}}in, G.~K. Kurt, and
  C.~Ke{\c{c}}eci, ``Robust and fast automatic modulation classification with
  cnn under multipath fading channels,'' in \emph{2020 IEEE 91st Vehicular
  Technology Conference (VTC2020-Spring)}.\hskip 1em plus 0.5em minus
  0.4em\relax IEEE, 2020, pp. 1--6.

\bibitem{44}
J.~Xu, C.~Luo, G.~Parr, and Y.~Luo, ``A spatiotemporal multi-channel learning
  framework for automatic modulation recognition,'' \emph{IEEE Wireless
  Communications Letters}, vol.~9, no.~10, pp. 1629--1632, 2020.

\bibitem{45}
F.~Zhang, C.~Luo, J.~Xu, and Y.~Luo, ``An efficient deep learning model for
  automatic modulation recognition based on parameter estimation and
  transformation,'' \emph{IEEE Communications Letters}, vol.~25, no.~10, pp.
  3287--3290, 2021.

\bibitem{46}
J.~Zhang, T.~Wang, Z.~Feng, and S.~Yang, ``Amc-net: An effective network for
  automatic modulation classification,'' in \emph{ICASSP 2023-2023 IEEE
  International Conference on Acoustics, Speech and Signal Processing
  (ICASSP)}.\hskip 1em plus 0.5em minus 0.4em\relax IEEE, 2023, pp. 1--5.

\bibitem{47}
Y.~Chen, B.~Dong, C.~Liu, W.~Xiong, and S.~Li, ``Abandon locality: Frame-wise
  embedding aided transformer for automatic modulation recognition,''
  \emph{IEEE Communications Letters}, vol.~27, no.~1, pp. 327--331, 2022.

\end{thebibliography}
\end{document}